\documentclass [prl, twocolumn, superscriptaddress, floatfix]{revtex4}
\usepackage{graphicx}
\usepackage{dcolumn}
\usepackage{bm}
\usepackage{amsmath,amssymb,amsfonts,latexsym}
\usepackage{times}
\def\unit#1{\ensuremath{\mathrm{\,#1}}}
\newcommand{\br}{{\bf r}}
\newcommand{\bx}{{\bf x}}

\begin{document}

\title
 {What buoyancy really is. A Generalized Archimedes Principle\\ for sedimentation and ultracentrifugation}

\author{Roberto Piazza}
\affiliation{Department of Chemistry (CMIC), Politecnico di
Milano, via Ponzio 34/3, 20133 Milano, Italy}
\email{roberto.piazza@polimi.it}%
\author{Stefano Buzzaccaro}
\affiliation{Department of Chemistry (CMIC), Politecnico di
Milano, via Ponzio 34/3, 20133 Milano, Italy}
\author{Eleonora Secchi}
\affiliation{Department of Chemistry (CMIC), Politecnico di
Milano, via Ponzio 34/3, 20133 Milano, Italy}
\author{Alberto Parola}
\affiliation{Department of Science and High Technology,
Universit\`a dell'Insubria, Via Valleggio 11, 22100 Como, Italy}

\begin{abstract}
Particle settling is a pervasive process in nature, and
centrifugation is a much versatile separation technique. Yet, the
results of settling and ultracentrifugation experiments often
appear to contradict the very law on which they are based:
Archimedes Principle - arguably, the oldest Physical Law. The
purpose of this paper is delving at the very roots of the concept
of buoyancy by means  of a combined experimental-theoretical study
on sedimentation profiles in colloidal mixtures. Our analysis
shows that the standard Archimedes' principle is only a limiting
approximation, valid for mesoscopic particles settling in a
molecular fluid, and we provide a general expression for the
actual buoyancy force. This ``Generalized Archimedes Principle''
accounts for unexpected effects, such as denser particles floating
on top of a lighter fluid, which in fact we observe in our
experiments.
\end{abstract}
\maketitle

Sedimentation of particulate matter is ubiquitous in the natural
environment and widespread in industrial processes. For instance,
particle and biomass settling is responsible for the formation of
depositional landforms~\cite{Julien} and plays a crucial role in
marine ecology~\cite{Kiorboe}, while centrifugation of insoluble
solids is a valuable separation methods in the extractive,
chemical, and food processing industry~\cite{Leung}. Thanks to the
genius of Jean Perrin, sedimentation studies also provided the key
support to the theory of Brownian motion~\cite{Perrin}, and
originated powerful methods to investigate soft and biological
matter, such as ultracentrifugation, a standard tool to obtain the
size distribution of biological macromolecules or to pellet
cellular organelles and viruses~\cite{Lebowitz}. A  particle
settling in a simple fluid is subjected, besides to its weight, to
an upward buoyancy force that, according to Archimedes' principle,
is given by the weight of the displaced fluid. Usually, however,
the settling process involves several dispersed species, either
because natural and industrial colloids display a large size
distribution, or because additives are put in on purpose. The
latter is the case of density--gradient ultracentrifugation (DGU),
where heavy salts, compounds like iodixanol, or more recently
colloidal nanoparticles, are added to create a density gradient in
the solvent. In DGU, proteins, nucleic acids, or cellular
organelles are expected to accumulate in a thin band around the
position in the cell where the local solvent density matches the
density of the fractionated species, the so-called isopycnic
point.

DGU is extremely sensitive, allowing for instance to resolve
differently labeled genomes with high efficiency~\cite{Lueders},
yet a subtle puzzle recurs in several studies. Even in earlier DGU
measurements, the apparent density of some proteins was found to
depend on the medium used to establish the density
gradient~\cite{Ifft}. The advent of sol--based DGU, allowing not
only for more efficient separation of
cells~\cite{Pertoft,Claassens}, but also for fractionation of
carbon nanotubes~\cite{Bonaccorso} and graphene~\cite{Green},
brought out more striking discrepancies. Indeed, the isopycnic
densities of organelles~\cite{Pertoft} or carbon
nanotubes~\cite{Bonaccorso} fractionated using Percoll$^{TM}$, a
standard DGU sol, are markedly different from those found in
sucrose or salt gradients, and striking anomalies have been
observed even for simple polystyrene latex
particles~\cite{Morganthaler}. What value should then we take for
the density of the medium, to predict the isopycnic point, if the
surrounding fluid is not a simple liquid, but rather a complex
mixture including other particulate species of different size
and/or density? Similar ambiguities exist in experimental and
numerical studies of colloid mixture settling in fluidized
beds~\cite{Gibilaro,Wielen}, where it is highly debated whether
the density $\rho$ of the bare \emph{solvent}, or rather the
density $\rho_s$ of the \emph{suspension} should be used to
evaluate the buoyant force. The latter choice is more widespread,
but both attitudes have been taken in the
literature~\cite{Poletto}, and even empirical interpolating
expressions have been suggested to fit experimental
data~\cite{Ruzicka,Grbavcic}.
\begin{figure}
\includegraphics [width =\columnwidth]{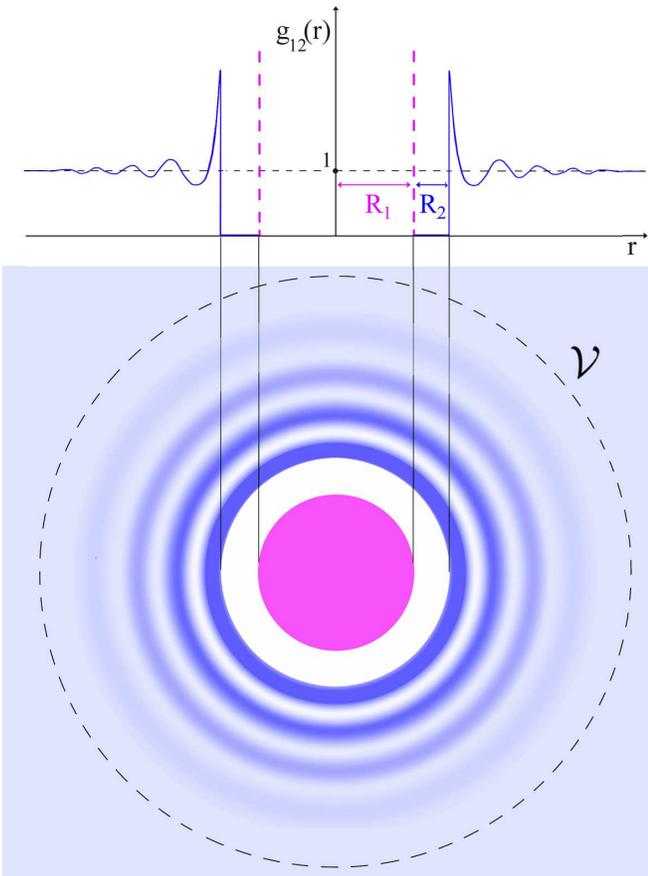}
\caption{\label{f1} \emph{(Color on line)} Schematic view of the
density perturbation induced in the surrounding fluid by a
settling colloidal particle or radius $R_1$, with the upper panel
showing the mutual radial correlation function $g_{12}(r)$ of
type-2 particles or radius $R_2 = qR_1$. The small $q$ (or
low-density) approximation leading to Eq.~(\ref{isopycnic})
corresponds to evaluate density changes by just taking into
account the white ``depleted'' spherical shell lying between $R_1$
and $R_2$.}
\end {figure}

The key point of our argumentation is that, when the suspending
fluid is a colloidal suspension or a highly structured solvent,
the amount of ``displaced fluid'' occurring in the simple
Archimedes' expression is substantially modified by the density
perturbation induced by the particle itself in the surrounding. We
shall focus on binary mixtures of particles of type 1 and 2, whose
volumes and material densities are respectively given by $(V_1,
\rho_1)$ and $(V_2, \rho_2)$, suspended in a solvent of density
$\rho$, under the assumption that component 1 is very diluted. Let
us consider, as in Fig.~\ref{f1}, a large spherical cavity of
volume $\mathcal{V}$ surrounding a single type-1 particle, and try
to extend the common argument used to derive the Archimedes'
principle. In the absence of particle 1, mechanical equilibrium
requires the total pressure force exerted by the external fluid on
$\mathcal{V}$ to balance exactly the weight $W=
m_2n_2g\mathcal{V}$, where $n_2$ is the number density of type-2
particles and $m_2 = (\rho_2 -\rho)V_2$ their buoyant mass. When
particle 1 is inserted, however, the distribution of type-2
particles in $\mathcal{V}$ changes, because interactions generate
a concentration profile set by the mutual radial distribution
function $g_{12}(r)$, which quantifies the local deviations from
uniform density~\cite{Hansen}. The total weight of the type-2
particles in $\mathcal{V}$ is now given by $W' =
m_2gn_2\int_\mathcal{V}g_{12}(r)\mathrm{d}^3r$. By taking the size
of the cavity much larger than the range of $g_{12}(r)$, the total
mass contained in $\mathcal{V}$ will then be subjected to an
unbalanced mechanical force~\footnote{We assume that the both
mutual interactions between the two species and self interactions
between type-2 particles are sufficiently short--ranged.
Eq.~(\ref{GAP}) is also valid when the host suspension is non
uniform, provided that $n_2$ varies slowly over the range of
$g_{12}(r)$.}

\begin{equation}\label{GAP}
    F_1 = W - W' = - m_2g n_2\int \left[g_{12}(r)-1\right]
    \mathrm{d}^3r.
\end{equation}
Provided that the density correlations embodied by $g_{12}(r)$ are
fully established, $F_1$ will also amount to an effective
\emph{excess buoyancy force} acting on the test particle, which
adds up to the usual Archimedes' term $F_0 = -\rho V_1g$. This
``Generalized Archimedes Principle'' (GAP), which is our main
theoretical result, can be equivalently written in terms of purely
thermodynamic quantities. Provided that the number density $n_1$
of type-1 particles is very low, it is indeed easy to show that
(see Supplementary Material):
\begin{equation}\label{thermoGAP}
   F_1 = m_2 g\left ( \frac{\partial
\Pi}{\partial n_2} \right )^{-1}\left[\frac{\partial \Pi}{\partial
n_1} -k_BT\right],
\end{equation}
where $\Pi$ is the osmotic pressure of the suspension.
Eq.~(\ref{thermoGAP}) shows that $F_1$ is proportional to the
buoyant mass of type-2 particles and to the osmotic
compressibility, whereas the last factor explicitly accounts for
mutual interactions between the two components.

For spherical particles of radii $R_1$ and $R_2$, a simple
expression for $F_1$ can be derived provided that component 2 is
very diluted too, or, alternatively, that the range of $g_{12}(r)$
is much smaller than $R_1$, which is usually the case if  the size
ratio $q= R_2/R_1 \ll 1$. In this limit, taking $ g_{12}(r) = 0$
for $r<R_1+R_2$, and 1 otherwise, we get $F_1 = (4\pi/3)
(R_1+R_2)^3 n_2m_2g$.  This result has a simple physical
explanation: the excess buoyancy comes from the type-2 particle
excluded from the depletion region shown in white in Fig.1. The
total buoyancy $F_1+F_0$ yields an ``effective'' density of the
suspending fluid
\begin{equation}\label{effdens}
    \rho^* =  \rho + \Phi_2 (1+q)^3 (\rho_2-\rho),
\end{equation}
where $\Phi_2$ is the volume fraction of type-2 particles. Note
that, assuming $\rho_2>\rho$, $\rho^*$ is always larger than
\emph{both} $\rho$ and $\rho_s = \rho + (\rho_2-\rho)\Phi_2$.
Hence, the empirical interpolating expression suggested
in~\cite{Ruzicka} is incorrect. A straightforward consequence is
that the weight of a type-1 particle is exactly balanced by a
suspension of type-2 particles at volume fraction:
\begin{equation}\label{isopycnic}
    \Phi_2^* = \frac{\Phi_2^{iso}}{(1+q)^3},
\end{equation}
which can be substantially \emph{lower} than the isopycnic value
\mbox{$\Phi_2^{iso}= (\rho_1-\rho)/(\rho_2-\rho)$} one would get
from assuming $\rho^*$ equal to the suspension density. In the
general, however, the additional force $F_1$ may not necessarily
oppose gravity. A strong attractive contribution to the mutual
interaction  may indeed overbalance the excluded volume term we
considered, reversing the sign of $F_1$. Hence, particle 1 can
actually be \emph{pulled down} by the surrounding, showing an
apparently larger density.

Although derived for colloid mixtures, Eq.~(\ref{GAP}) is valid in
much wider conditions, whenever the region of perturbed solvent
density is not negligible compared to $V_1$. Moreover, being
solely based on a force balance argument, Eq.~(\ref{GAP}) does
\emph{not} require the suspension to have reached sedimentation
equilibrium, but only that the density distribution of type-2
particles around particle 1 has fully settled.  Hence, since the
time scale for the latter is usually much faster (at least for
Brownian particles), these predictions could be in principle
checked on settling mixtures or in fluidized bed experiments. In
practice, however, telling apart buoyancy effects from viscous
forces is quite hard, because of the presence of long--range
hydrodynamic interactions~\cite{Buscall}.

Thus, to test these ideas, we have devised a targeted
\emph{equilibrium} measurement. We have studied model colloidal
mixtures, obtained by adding a minute quantity ($\Phi_1 \le
10^{-5}$) of polymethyl-methacrylate (PMMA, $\rho_1 =
1.19\unit{g/cm^3}$, obtained from microParticles GmbH, Berlin)
particles with three different particle sizes \mbox{($R_1\simeq
220, 300, 400\unit{nm}$)}, to a moderately concentrated suspension
of spherical particles with radius $R_2 = 90\unit{nm}$ made of
MFA, a tetrafluoroethylene copolymer with density $\rho_2 =
2.14\unit{g/cm^3}$~\cite{Degiorgio}. MFA particles, though
spherical and monodisperse, are partially crystalline, and
therefore birefringent. Their intrinsic optical anisotropy yields
a depolarized component $I_{VH}$ in the scattered light that does
\emph{not} depend on interparticle interactions, but only on the
local particle concentration~\cite{Degiorgio}. Hence, the full
equilibrium sedimentation profile can be simply obtained by
vertically scanning a mildly focused laser beam and measuring
$I_{VH}$ as a function of the distance from the cell bottom. A
simple numerical integration of the experimental profile yields
moreover the full equation of state of the system~\cite{Piazza,
Buzzaccaro}. In addition, MFA has a very low refractive index $n =
1.352$, so it scatters very weakly in aqueous solvents. For better
index--matching, we have used as solvent a solution of urea in
water at 15\% by weight, with density $\rho =1.04\unit{g/cm^3}$.
Hence, at equilibrium, the PMMA particles can be visually spotted
as a thin whitish layer lying within a clear MFA sediment.

\begin{figure}
\includegraphics [width =\columnwidth]{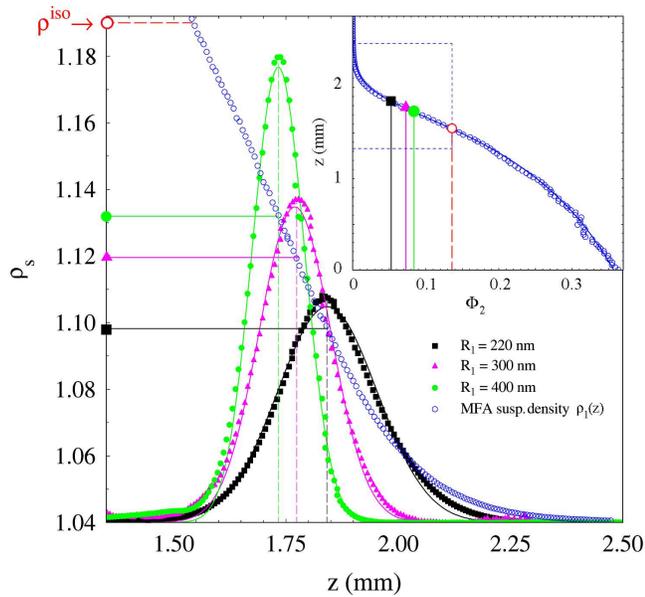}
\caption{\label{f2} (\emph{Color on line}) Inset: Equilibrium
sedimentation profile of a suspension of MFA particles with radius
\mbox{$R_2 =90\unit{nm}$}, dispersed in a solution of urea in
water with density \mbox{$\rho =1.04\unit{g/cm^3}$}. Here $z$ is
the distance from the cell bottom, $\Phi_2(z)$ the local MFA
volume fraction, and the full line is the theoretical profile for
hard-spheres with a radius $R'\simeq 1.1 R_2$. On the profile, the
mean position of the thin layers of PMMA particles with radius 400
(bullet), 300 (triangle), and $220\unit{nm}$ (square) are compared
to the prediction from the simple Archimedes' principle (open dot,
corresponding to $\Phi_2 = 0.136$). Main body: Expanded view of
the profile region within the rectangular box in the inset,
showing the local density $\rho_s$ of the MFA suspension.
Superimposed are the full distributions (with normalized area) of
the PMMA particles obtained from turbidity measurements and fitted
with gaussian distributions as described in the text. Note the
location of the isopycnic point where $\rho_s = \rho_2$.}
\end {figure}

The equilibrium sedimentation profile of the MFA suspension
obtained by DeLS, is shown in the inset of Fig.~\ref{f2}. Using
the simple Archimedes' principle, we would expect the PMMA
particles to gather around the isopycnic level, namely, the region
where the local suspension density is about $1.19\unit{g/cm^3}$,
which corresponds to $\Phi_2^{iso} = 0.136$. However, the layers
lie well above this level, the more the smaller the PMMA particles
are. The distribution of the guest particles can be obtained by
evaluating via turbidity measurements the sample extinction
coefficient through the layer, where the PMMA peak concentration
does not exceed $\Phi_1 \simeq 10^{-4}$. The body of Fig.~\ref{f2}
shows that the normalized probability distributions for the PMMA
particle position have a bell shape centered on anomalously high
$z$-values, with a width that grows with decreasing PMMA particle
size. Since the MFA profile changes very smoothly on the scale of
the layer thickness, it is in fact easy to show (see Supplementary
Material) that the PMMA particles should approximately distribute
as a gaussian with standard deviation:
\begin{equation}\label{PMMA_band}
\sigma \simeq
\sqrt{\Phi_2^*\,\ell_{g1}\left|\frac{\mathrm{d}\Phi_2}{\mathrm{d}z}\right|_{z=z^*}^{-1}},
\end{equation}
where $\ell_{g1} = k_BT/m_1g$ is the gravitational length of the
type-1 particles, which we assume to be much larger than $R_1$ and
$R_2$. Table 1 shows that the experimental values agree very well
with the values predicted by Eq.~(\ref{isopycnic}), both for the
effective isopycnic point $\Phi_2^*$ and for the standard
deviations of the gaussian fits.

\begin{table}
\caption{\label{t1}Theoretical and experimental values for the
effective isopycnic points $\Phi_2^*$ and for the standard
deviation of the gaussian fits to the PMMA profiles. Calculated
values are based on the simple ``excluded volume'' approximation
leading to Eq.~(\ref{isopycnic}) and~(\ref{PMMA_band}), which may
be reasonably expected to hold because the values of $\Phi_2^*$
are rather small and $q$ not too large.}
\begin{ruledtabular}
\begin{tabular}{ccccccc}
$R_1$ (nm) & $q$ & $\ell_{g1}\unit{(\mu m)}$&$\Phi_2^{*\unit{theo}}$ & $\Phi_2^{*\unit{exp}}$  & $\sigma^{\mathrm{teo}}\unit{(\mu m)}$ & $\sigma^{\mathrm{exp}} \unit{(\mu m)}$ \\

  220 & 0.41 & 63& 0.049 & 0.052 & 110 & 113 \\
  300 & 0.30 & 24& 0.062 & 0.072 & 78 & 80 \\
  400 & 0.22 & 10& 0.074 & 0.083 & 55 & 58 \\
\end{tabular}
\end{ruledtabular}
\end{table}

\begin{figure}
\includegraphics [width =\columnwidth]{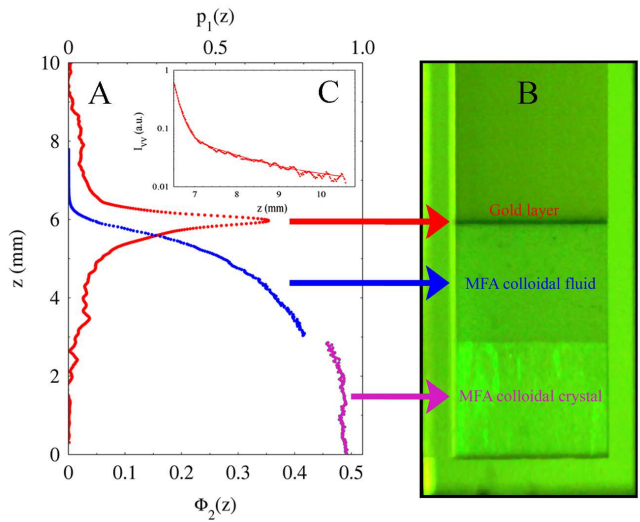}
\caption{\label{f3} (\emph{Color on line}) Equilibrium
sedimentation profile (A) and visual appearance (B) of a MFA
suspension with a little amount of $R_1 \simeq 16$ nm gold
particles added. As evidenced by the weak Bragg reflections, the
phase closer to the cell bottom is a colloidal crystal, whereas
the upper phase is a colloidal fluid. To enhance the visibility of
the thin gold layer, the sample picture has been taken using green
illumination, with a narrow wave-length band around the plasmonic
absorbtion peak of the gold colloid. The concentration profile
obtained from turbidity data (exploiting in this case the
proportionality between gold \emph{absorption} and local
concentration) shows that gold particles are also present both
within the MFA sediment and in the supernatant solvent. The
semilog plot of the polarized scattering intensity in Panel C is
fitted with a double exponential, as discussed in the text.}
\end {figure}

When considering the opposite case of small, dense particles
settling in a ``sea'' of larger but lighter ones, the GAP yields
rather surprising predictions. Eq.~(\ref{thermoGAP}) shows indeed
that $F_1$ is proportional to the weight of a \emph{large}
particle: actually, the density perturbations in the host
suspension can generate an excess buoyant force $F_1$ amounting to
a sizable fraction of $m_1g$, thus yielding an upward push on the
small particle that largely \emph{outbalances} its own weight.
More specifically, in the Supplementary Material we show that, for
hard-sphere mixtures with $q\gg1$, $F_1$ is strongly
non-monotonic, reaching a maximum at $\Phi_2\lesssim 0.2$. Hence,
most of the denser particles will accumulate \emph{atop} the
lighter ones~\footnote{Note that the accumulation on top of the
heavier particles does \emph{not} however lead to a
macroscopically inverted density profile. An attentive examination
of Eq.~(\ref{GAP}) shows indeed that the weight increase with
respect to a suspension of type-2 particles at volume fraction
$\Phi^*$, due to the presence of the heavier particles, is exactly
balanced by the ``expulsion'' from the accumulation layer of those
particles of type 2 that yield the excess buoyancy $F_1$.
Macroscopic hydrodynamic stability is thus preserved.}. A striking
example of this rather weird effect is shown in Fig.~\ref{f3},
where gold particles, with a radius of about $16\unit{nm}$ and a
density $\rho_1\simeq 19.3\unit{g/cm^3}$ are seen to float mostly
in the upper, very dilute region of an equilibrium sedimentation
of MFA particles (here $q \simeq 5.6$). The DeLS profile shows
that the MFA suspension is actually a colloidal \emph{fluid} (not
a solid), with a density as low as $\rho_s \simeq
1.2\unit{g/cm^3}$ around the region where most of the gold
particles accumulate. Since, for $\Phi_2\rightarrow 0$, the excess
buoyant force $F_1$ vanishes, some of the latter must lie within
the MFA fluid phase too with a concentration profile that
decreases downwards, as confirmed by turbidity data. Similarly,
gold particles are expected to distribute in the supernatant
solvent too, according to a barometric law $c(z) \propto
\exp(-z/\ell_{g1})$, with a gravitational length $\ell_{g1}\simeq
1.4\unit{mm}$. This weak barometric region can be detected by
polarized light scattering~\footnote{Although in index--matching,
MFA particles still scatter polarized light, which is however
fully incoherent and proportional to the depolarized scattered
intensity~\cite{Degiorgio}.}. Panel C in Fig.~\ref{f3} shows that
the polarized scattered intensity can be fitted as the weighted
sum of two exponentials $I = I_1\exp(-z/\ell_{g1})+ I_2
\exp(-z/\ell_{g2}),$ where the MFA gravitational length is fixed
at the value $\ell_{g2} =0.13\unit{mm}$, whereas from the fit
$\ell_{g1} \simeq 1.38\unit{mm}$ for gold. This value for
$\ell_{g1}$ corresponds to an average particle radius $R_1 \simeq
16\unit{nm}$, in very good agreement with the estimate made from
the position of the particle plasmonic absorption peak at $\lambda
= 528\unit{nm}$.

The GAP qualitatively accounts for the anomalous DGU measurements
of polystyrene bead density~\cite{Morganthaler}, even when, in the
presence of oppositely--charged nanoparticles, the latter
apparently \emph{increases}, and for empirical expressions used to
fit flotation--bed experiments~\cite{Ruzicka, Grbavcic}. But
Eq.~(\ref{GAP}) has a much wider scope.  For instance, provided
that a model for $g_{12}$ is available, it should correctly
account for ``solvation'' effects on the buoyancy force felt by
proteins, simple molecules, even single ions, or provide a
sensitive way to detect by DGU aggregation and association effects
in biological fluids. Similarly, corrections to the simple
Archimedes' expression will also show up for nanoparticles
settling in a strongly correlated solvent, such as a pure fluid or
a liquid mixture close to a critical point. Some relation with the
Brazil nut effect in granular fluids, which is also affected by
the densities of the grain~\cite{Mobius, Kudrolli}, may also
exist, although the latter is usually complicated by the presence
of dissipation, convective effects, and effective thermal
inhomogeneity. In fact, due to its exquisite sensitivity to the
specific properties of a mixture, the ``reversed''
gravity--segregation effect we have highlighted may allow to
devise novel sophisticated DGU fractionation methods, able to tell
apart solutes with the same density and composition, but different
size.
\begin{acknowledgments}
We thank D. Frenkel, P. Chaikin, M. Dijkstra, H. Stone, F.
Sciortino, R. Stocker, R. van Roij, A. Philipse, A. van Blaaderen,
R. Golestanian,
 D. Aarts,  C. Likos, L. Cipelletti, L. Berthier, L. Isa, P. Cicuta, and V. Degiorgio for a critical reading of the manuscript, and Solvay Specialty Polymers (Bollate, Italy)
 for the kind donation of the MFA sample batch.  This work was supported by the Italian Ministry of Education and Research (MIUR - PRIN Project
2008CX7WYL).
\end{acknowledgments}

\section{Supplementary Material}
\small
\subsection{Effective buoyancy.} We provide here a formal
derivation of the buoyancy force $F_1$ acting onto a test type-$1$
colloid immersed in a solution of type-$2$ particles, expressed in
purely thermodynamic terms. The density profile of a suspension of
particles in the presence of a gravitational field is described by
the hydrostatic equilibrium condition
\begin{equation}\label{hydro}
\frac{d \Pi[n_2(z),T]}{dz}= - m_2g\, n_2 (z), \tag{S1}
\end{equation}
where $m_2$ is the buoyant mass of type-2 particles, $\Pi$ the
osmotic pressure, and we assume that the number density $n_2$ may
depend on $z$. The gravitational length $\ell_g=k_BT/(m_2g)$
defines the characteristic scale of the spatial modulations of the
density profile: here and in the following we assume that $\ell_g$
is the largest length in the problem, a condition easily met in
colloidal suspensions. Under this assumption, the contribution to
the buoyancy force acting onto a test particle (denoted by index
$1$) inserted in this solution, due to the presence of type-$2$
particles, is given by Eq. (1) in the text:
\begin{equation}
F_1(z)=-m_2\, g\,n_2(z)\,\int d\br\,h_{12}(r), \tag{S2}
\label{buo}
\end{equation}
where $h_{12}(r) = g_{12}(r)-1$. This expression depends on the
mutual correlations between the two species but can be
equivalently written in terms of purely thermodynamic quantities.
Regarding the system as a binary mixture where component $1$ is
extremely diluted, the Ornstein-Zernike relation in the $n_1\to 0$
limit (see Ref. [14])
\begin{equation}
h_{12}(r) = c_{12}(r) + n_2 \, \int d\bx c_{12}(\br-\bx)
\,h_{22}(\bx) \tag{S3}
\end{equation}
allows to express the integral of $h_{12}(r)$ in terms of the
integral of the direct correlation function $c_{12}(r)$ and the
long wave-length limit of the structure factor of a type-$2$ one
component fluid $S_{22}(0)$:
\begin{equation}
\int d\br\,h_{12}(r) = S_{22}(0)\,\int d\br\,c_{12}(r). \tag{S4}
\end{equation}
Both terms at right hand side can be expressed as thermodynamic
derivatives of the Helmholtz free energy of the mixture $A$ via
the compressibility sum rules (Ref. [17]):
\begin{align*}
n_2\,S(0) =& k_BT\,\left [\frac{\partial^2 \left(A/V\right)}{\partial n_2^2} \right ]^{-1} \notag \\
k_BT\,\int d\br\,c_{12}(r) =& -\frac{\partial^2
\left(A/V\right)}{\partial n_1\,\partial n_2}. \tag{S5} \label{oz}
\end{align*}
According to the McMillan-Mayer theory of solutions, the
contribution of the solvent to the total free energy can be
disregarded if effective interactions among particles are
introduced. In the limit $n_1\to 0$ we can express the free energy
derivatives appearing in Eq. (\ref{oz}) in terms of the osmotic
pressure:
\begin{equation}
\Pi = -\frac{A}{V} + n_2\,\frac{\partial\left(A/V\right)}{\partial
n_2} + n_1\,\frac{\partial\left(A/V\right)}{\partial n_1} \tag{S6}
\end{equation}
leading to
\begin{align*}
F_1=&\frac{\partial^2\left(A/V\right)}{\partial n_1\partial n_2}
\,\left [ \frac{\partial^2\left(A/V\right)}{\partial n_2^2}\right
]^{-1}
\,m_2 g \\
=& \left [ \frac{\partial \Pi}{\partial n_1} -k_BT \right ]
\,\left [ \frac{\partial \Pi}{\partial n_2} \right ]^{-1} \,m_2 g
\label{buoy} \tag{S7}
\end{align*}
which coincides with Eq.~(2) in the paper. This shows that the
contribution to the buoyancy force on a type-$1$ particle due to
the presence of component $2$ is proportional to the buoyant mass
$m_2$. It is interesting to investigate the limiting form of the
buoyancy force when the type-$1$ particle is just a ``tagged"
type-$2$ particle, with identical physical properties. In this
case the system is effectively one-component and then
$\frac{\partial \Pi}{\partial n_1}=\frac{\partial \Pi}{\partial
n_2}$. The buoyancy force acting onto a particle in the solution
acquires the form:
\begin{equation}\label{singlecomp}
F = mg\left [ 1 - k_BT\left(\frac{\partial \Pi}{\partial
n}\right)^{-1}\right ] \tag{S8}.
\end{equation}

It is instructive to deduce Eq.~(\ref{singlecomp}) with a
different approach, which highlights its physical meaning. The
equilibrium sedimentation profile of a suspension of interacting
Brownian particles is usually derived by balancing gravity with
the diffusive term deriving from gradients in the osmotic
pressure. However, fixing the attention on a single test particle,
we can try to summarize the effect of all the other particles as
an ``effective field'' $F$ adding to the bare gravitational force
$-mg$. From the Smoluchowski equation, the combination of these
two contributions yield a density profile:
\begin{equation*}
k_BT {d n \over dz}=n (F-mg) , \label{profile}
\end{equation*}
that, combined with the hydrostatic equilibrium
equation~(\ref{hydro}), yields for $F$ the expression in
Eq.~(\ref{singlecomp}). Hence, the equilibrium sedimentation
profile of an interacting suspension can be equivalently viewed in
terms of the probability distribution for the position of a test
particle subjected to a spatially--varying gravitational field,
whose dependence on $z$ is dictated by the equation of state of
the suspension.

\begin{figure}[h]
\centering
\includegraphics[clip, width=\columnwidth]{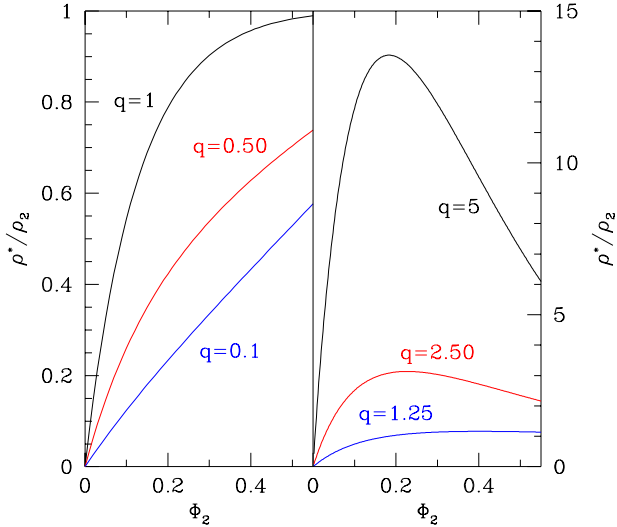}
\caption{\label{effmass}Effective mass density of the surrounding
medium, relative to the type-$2$ mass density as a function of
$\Phi_2$ for different $q=R_2/R_1$. Left panel: results for $q \le
1$. Right panel: $q > 1$. Note the change of scale in the vertical
axis. }
\end{figure}

In hard sphere systems we can easily obtain an approximate
expression for the buoyancy force from Eq. (\ref{buoy}): a rough
estimate of the excluded volume effects in the osmotic pressure
can be obtained following the familiar Van der Waals argument:
\begin{align*}
\Pi(n_1,n_2)-n_1k_BT &=\frac{N_2\,k_BT}{V-N_1\frac{4}{3}\pi (R_1+R_2)^3} \\
&\sim n_2 k_BT\,\left [ 1 + n_1\,\frac{4}{3}\pi (R_1+R_2)^3\right
] \tag{S9}
\end{align*}
By substituting this form into Eq. (\ref{buoy}) we recover the
simple result, already quoted in a slightly different form in the
main paper
\begin{equation}\label{force_simple}
F_1 = m_2 g \,\Phi_2\,\left(1+\frac{1}{q}\right)^3 \tag{S10}
\end{equation}
A more careful evaluation is obtained by starting from the
analytical expression of the excess free energy of a binary hard
sphere mixtures provided by Mansoori \emph{et al.} (J. Chem. Phys.
{\bf 54}, 1523, 1971). The result can be conveniently expressed in
terms of the {\sl effective mass density of the surrounding
medium} $\rho^*$ defined by
\begin{equation}
F_1 = \frac{4}{3}\pi\, R_1^3 \rho^* g \label{rhot} \tag{S11}
\end{equation}
The explicit expression for the effective density reads:
\begin{widetext}
\begin{equation*}
\frac{\rho^*}{m_2 n_2} =\frac{6 +(1-q)^2(2+q)\,(1-\Phi_2)^3 -
3(1-q^2)(1-\Phi_2)^2 - 2\,\left [(1-q)^2(2+q)-q^3\right
]\,(1-\Phi_2)}{(1-\Phi_2)^4 +\Phi_2(8-2\Phi_2)}
\end{equation*}
\end {widetext}
The dependence of $\rho^*$ on the size and volume fraction of
type-$2$ particles is shown in Fig.~\ref{effmass}. For $q>1$, i.e.
when a small test particle is immersed into a suspension of big
particles, the buoyancy force displays a pronounced maximum. In
the $q\to\infty$ limit, the maximum buoyancy force is attained at
$\Phi_2 \sim 0.154$, where it reduces to a sizeable fraction of
the effective weight of a type-$2$ particle: $F_1 \sim 0.055\,m_2
g$.

\subsection{Distribution of guest particles at equilibrium.} The
hydrostatic equilibrium condition for a suspension of type-$1$
particles reads:
\begin{equation}
\frac{d\Pi}{dz} = n_1\,\left [-m_1\,g +F_1\right] \tag{S12}
\label{hyd}
\end{equation}
where $\Pi$, $n_1$ and $m_1$ are the osmotic pressure, average
local density and buoyant mass respectively. In the limit of short
range interspecies correlations, the excess buoyant force $F_1$
due to the presence of type-$2$ particles is given by
Eq.~(\ref{force_simple}), while in the diluted limit of type-$1$
particles the ideal gas equation of state \mbox{$\Pi_1 = n_1
k_BT$} holds. Substituting these results in Eq. (\ref{hyd}) we
find:
\begin{equation}
k_BT\,\frac{d n_1}{dz} = n_1\,g\,\left [-m_1+m_2\,\Phi_2(z) \left
( 1+\frac{1}{q}\right )^3\right] \label{prof} \tag{S13}
\end{equation}
which defines the number density profile of type-$1$ particles.
The maximum of the resulting distribution corresponds to the
vanishing of the right hand side of this expression, given by
condition (3) of the main paper:
\begin{equation}
\Phi_2^*\equiv\,\Phi_2(z^*) = \frac{\Phi_2^{iso}}{(1+q)^3}
\tag{S14}
\end{equation}
where $\Phi_2^{iso} = (m_1/m_2) q^3$ coincides with the isopycnic
volume fraction defined in the main paper. By expanding
$\Phi_2(z)$ around the position of this maximum $z^*$, Eq.
(\ref{prof}) becomes:
\begin{align*}
\frac{d n_1}{dz} = &
n_1\,\frac{m_2\,g}{k_BT}\,\frac{d\Phi_2(z)}{dz}\Big\vert_{z=z^*}
\left ( 1+\frac{1}{q}\right )^3 (z-z^*) \\
= &
n_1\,\frac{d\Phi_2(z)}{dz}\Big\vert_{z=z^*}\,\frac{(z-z^*)}{\ell_{g1}\,\Phi_2^*}
\tag{S14} \label{prof2}
\end{align*}
whose solution $ n_1(z)$ is a gaussian centered in $z=z^*$ with
standard deviation given by Eq.~(5) of the main paper.

\end{document}